\documentclass[10pt,twocolumn,aps,prb,superscriptaddress]{revtex4-1}
\usepackage[latin9]{inputenc}
\usepackage{graphicx}
\begin{document}

\title{Generating indistinguishable photons from a quantum dot in a noisy
environment}

\author{Ted S. Santana}

\author{Yong Ma}
\altaffiliation{Current address: Chongqing Institute of Green and Intelligent Technology, Chinese Academy of Sciences, Chongqing 400714, China}

\author{Ralph N. E. Malein}

\affiliation{Institute of Photonics and Quantum Sciences, SUPA, Heriot-Watt University,
Edinburgh, United Kingdom}

\author{Faebian Bastiman}
\author{Edmund Clarke}

\affiliation{EPSRC National Centre for III-V Technologies, University of Sheffield,
United Kingdom}

\author{Brian D. Gerardot}
\email{b.d.gerardot@hw.ac.uk}

\affiliation{Institute of Photonics and Quantum Sciences, SUPA, Heriot-Watt University,
Edinburgh, United Kingdom}

\date{\today}
\begin{abstract}
Single photons from semiconductor quantum dots are promising resources for linear optical quantum computing, or, when coupled to spin states,
quantum repeaters. To realize such schemes, the photons must exhibit a high degree of indistinguishability. However, the solid-state environment
presents inherent obstacles for this requirement as intrinsic semiconductor fluctuations can destroy the photon indistinguishability. Here we use resonance fluorescence to generate indistinguishable photons from a single quantum dot in an environment filled with many charge-fluctuating traps. Over long time-scales ($>50$ $\mu$s), flickering of the emission due to significant spectral fluctuations reduce the count rates. Nevertheless, due to the specificity of resonance fluorescence, high-visibility two-photon interference is achieved.
\end{abstract}
\maketitle

\section{Introduction}

A semiconductor quantum dot (QD) can emulate a two-level or a multi-level atomic system. Two-level QD transitions can generate single photons
with a high degree of indistinguishability \citep{Patel_PRL_2008,Ates_PRL_2009,Gazzano_NatComms_2013,Atature_NatComms_2013,Voliotis_PRB_2014,Proux_PRL_2015,Thoma_PRL_2016,Somaschi_NatPhotonics_2016,Ding_PRL_2016,Dada_Optica_2016,Ralph_PRL_2016}, an ideal resource for implementing future quantum photonic technologies
such as Boson sampling \citep{Loredo_arxiv_2016,He_arxiv_2016} and linear optical quantum computing.
Multi-level QD systems, including the bi-exciton $\rightarrow$ exciton $\rightarrow$ ground-state cascade and so-called spin-$\lambda$
systems \citep{Xu_PRL_2008,Brunner_Science_2009} can be used to generate entangled photon pairs \citep{Akopian_PRL_2006,Salter_Nature_2010,Trotta_NatComm_2016,Chen_NatComm_2016} and spin-photon entanglement \citep{Gao_CLEO_2014,deGreve_Nat_2012,Schaibley_PRL_2013}, respectively, which can underpin implementations of quantum repeaters
and networks \citep{Northup_NatPhotRev_2014}. While a solid-state platform provides benefits for scalability \citep{Jons_NanoLett_2013},
functionality \citep{Warburton_Nature_2000,Trotta_AdvMat_2012}, and on-chip integration \citep{Luxmoore_PRL_2013,Kalliakos_APL_2014,Madsen_PRB_2014},
inherent fluctuations in the semiconductor matrix act as sources of noise \citep{Kuhlmann_NatPhys_2013,Matthiesen_SciRep_2014,Kuhlmann_Nat_2015,Stanley_PRB_2014} that inhomogeneously broaden a QD transition. The inhomogeneous broadening mechanisms can reduce the coherence (with a characteristic dephasing
time $T_{2}^{*}$) and indistinguishability (with a characteristic two-photon interference visibility $\nu(0)$) of photons extracted from the QD \citep{Thoma_PRL_2016,Legero_APB_2003}.

There are two primary inhomogeneous broadening mechanisms for semiconductor QDs at low temperature. Nuclear spin noise, caused by fluctuations
in the spin orientation of the QD constituent atoms' nuclei, results in a changing magnetic field that couples to an individual electron
or hole spin in the QD. In the context of a single photon source, nuclear spin noise can be mitigated by a few strategies. For the negatively
charged trion (X$^{1-}$), a modest external magnetic field in the growth direction (Faraday geometry) can screen the nuclear spin fluctuations
\citep{Ralph_PRL_2016} or a second laser can be used to partially stabilize the Overhauser field \citep{Kuhlmann_Nat_2015}. For the
neutral exciton (X$^{0}$), the effective magnetic field caused by the exchange interaction energy effectively mitigates the nuclear
spin noise \cite{Ralph_PRL_2016,Kuhlmann_Nat_2015}. The second, often more severe source of inhomogeneous broadening is localized charge
fluctuations in the QD environment which dynamically modify the electric field at the position of the QD and alter the dot's transition energy
via the Stark effect. Charge noise is present under both incoherent \citep{Robinson_PRB_2000,Berthelot_NatPhys_2006,Gazzano_NatComms_2013} and coherent \citep{Gerardot_APL_2011,Konthasinghe_PRB_2012,Houel_PRL_2012,Kuhlmann_NatPhys_2013,Hauck_PRB_2014,Matthiesen_SciRep_2014} excitation. The origins of the charge traps that host the fluctuations can vary depending on the sample; potential sources include nearby
surface states in processed photonic structures \citep{Wang_APL_2004,Majumdar_PRB_2011,Srinivasan_PRB_2014}, traps created at heterostructure interfaces \citep{Houel_PRL_2012}, impurities from intentional dopants \citep{Gerardot_APL_2011}, and
residual background dopant impurities \citep{Hauck_PRB_2014,Diederichs_PRB_2013}. Charge noise is often identified as an origin of increased ensemble
dephasing and decreased two-photon interference visibility \citep{Gazzano_NatComms_2013,Santori_Nature_2002,Thoma_PRL_2016,Madsen_PRB_2014,Voliotis_PRB_2014}.

Although dynamics in the local electric field cannot be universally eliminated in a semiconductor device, the impact of charge noise induced
inhomogeneous broadening on photon indistinguishability can be alleviated. One method is to reduce the emitter lifetime ($T_{1}$) so that it
masks $T_{2}^{*}$, typically achieved with Purcell enhancement by embedding the QD in a high-quality factor (Q) cavity \citep{Santori_Nature_2002,Laurent_APL_2005,Gazzano_NatComms_2013,Thomas_PRL_2015}. Unfortunately, increased charge noise often accompanies the Purcell enhancement in processed photonic devices \citep{Wang_APL_2004,Majumdar_PRB_2011,Srinivasan_PRB_2014}. Additionally, recapture of carriers excited by a non-resonant laser can ruin the purity of the single photon emission from QDs in cavities, especially when driven near saturation \citep{Strauf_NatPhot_2007,Gazzano_NatComms_2013,Madsen_PRB_2014}. An alternative method to suppress the effects of inhomogeneous broadening
is to exploit resonantly scattered photons, which offers two significant benefits with respect to $T_{2}^{*}$. Firstly, resonant excitation
eliminates populating charge traps via above band-gap excitation (although ambient charge fluctuations remain). Secondly, spectral fluctuations
occurring at time-scales greater than $T_{1}$ which detune the transition from the resonant laser by energy $\delta$ effectively increase the
fraction of elastically scattered photons at the expense of incoherent scattering \citep{Konthasinghe_PRB_2012}. Elastically scattered photons
are fundamentally indistinguishable \citep{Atature_NatComms_2013,Proux_PRL_2015}. Concurrently, the dynamic detuning of the two-level system from the
laser resonance leads to fluctuations in extracted photon rates. While this flickering has recently been characterized over a wide range of time-scales \citep{Kuhlmann_NatPhys_2013,Matthiesen_SciRep_2014,Kuhlmann_Nat_2015,Stanley_PRB_2014}, here we demonstrate that, in-spite of large intrinsic charge noise in a sample, highly indistinguishable single photons can be obtained. This result can be applied to emitters in less mature, emerging platforms
that thus far suffer from more spectral fluctuations than state-of-the-art III-V quantum dot samples. Examples include quantum dots emitting at telecom wavelengths \cite{Rima_APL_2016} or quantum emitters in two-dimensional materials, for which resonance fluorescence has recently been demonstrated \citep{Kumar_Optica_2016}.

\section{Experimental setup}

Our sample consists of self-assembled InGaAs QDs embedded in a GaAs Schottky diode for deterministic charge state control. The QDs are positioned at an antinode of a fifth-order planar cavity on top of a Au layer which functions as a mirror and Schottky gate \citep{Ma_JAP_2014}. The experiments are performed at $T=4$ K using confocal microscopy. Resonance fluorescence (RF) is separated from the reflected laser light using orthogonal linear polarizers in the excitation and collection arms of the confocal microscope \citep{Atature_NatComms_2013,Kuhlmann_NatPhys_2013,Matthiesen_SciRep_2014,Kuhlmann_Nat_2015}. The RF is detected with a silicon avalanche photodiode (jitter $\sim500$ ps) in photon counting mode and the arrival time of each photon is recorded. The RF is further characterized via high-resolution ($27$ MHz) spectroscopy using a Fabry-Perot interferometer ($5.5$ GHz free spectral range), a Hanbury Brown\textendash Twiss interferometer to measure second-order correlation functions, and an unbalanced Mach-Zender (MZ) interferometer ($t_{delay}=49.7$ ns) with polarization control
in each arm to measure postselected, two-photon interference.

\section{Results}

\subsection{Photoluminescence and photon statistics}

\begin{figure}[hb]
\centering \includegraphics[width=1\linewidth]{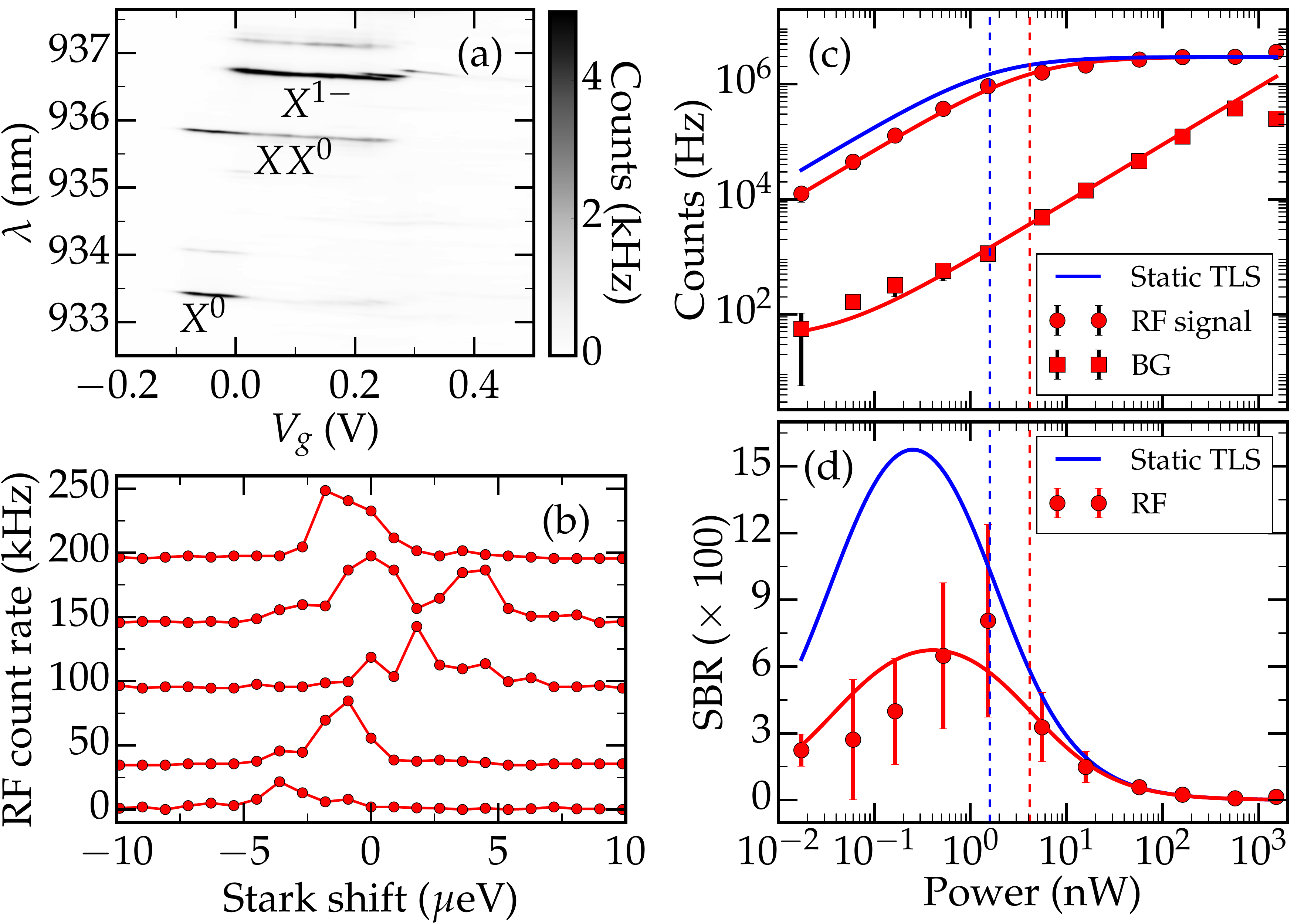} \caption{(a) PL map of an isolated charge tunable QD with the neutral exciton (X$^{0}$), the biexciton (XX$^{0}$) and the trion (X$^{1-}$) states labelled. (b) Five successive RF detuning spectra of the X$^{1-}$ transition with excitation power $P_{exc}=0.12(1)$~nW and $50$~ms integration time per data point. Due to charge noise, the transition energy fluctuates. (c) Saturation curve under strong spectral fluctuation (red circle) and the background due to back reflection of the driving field (red square). The expected saturation curve for a static TLS
is represented by the blue line. From the fit (solid red line), the measured saturation power is $P_{sat}=4.1(3)$~nW (dashed red line), while the hypothetical static TLS would have $P_{sat}=1.59$~nW (dashed blue line). (d) Measured signal to background ratio (SBR) under strong spectral fluctuation (red circles) compared with the SBR in the hypothetical case of no spectral fluctuation (blue line).}
\label{fig1} 
\end{figure}

We first characterize single QDs in the device using non-resonant excitation ($830$~nm), confirming bright photoluminescence and deterministic charging, as shown in Fig.~\ref{fig1}~(a). Separate measurements reveal that the lifetime of the X$^{1-}$ transition is $625$ ps, yielding a transform limited linewidth of $1.05$ $\mu$eV. Resonance between the driving field and the fluorescence of the QD is then established by setting the laser wavelength constant and sweeping the transition using applied voltage bias ($V_{g}$) across resonance. Fig.~\ref{fig1}~(b) shows five such RF spectra from the X$^{1-}$ transition measured
in rapid succession. Here, the excitation power is $P_{exc}=0.029(3)P_{sat}$, where $P_{sat}$ is the power at saturation, and the integration time per point is $50$~ms. Resonance is randomly achieved within an energy interval of about $10$ $\mu$eV due to significant spectral fluctuations. We ascribe this due to a large amount of charge noise in the QD environment that imposes its dynamics on the transition of the QD via the Stark shift. A saturation curve measuring the total number of detected photons ($I_{tot}$) as a function of excitation power is shown in Fig.~\ref{fig1}~(c). A fit to the curve (solid red line) using $I_{tot}\propto P_{exc}/[P_{exc}+P_{sat}]$, yields $P_{sat}=4.1(3)$ nW. The blue curve displays a curve for an ideal, static two-level system (TLS) \citep{Mollow_PR_1969} with
$T_{1}=0.625$ ns. The saturation power for this static TLS (blue dashed line) is $1.59$ nW, approximately three times smaller than the experimental value. Additionally, for excitation powers below saturation, $I_{tot}$ is reduced relative to the expected value for the static TLS. This impacts the signal to background ratio of the resonance fluorescence (Fig.~\ref{fig1}~(d)), where the background is the uncancelled reflected laser light. The relatively large error
bars are due to charge noise.

\begin{figure}[ht]
\centering \includegraphics[width=1\linewidth]{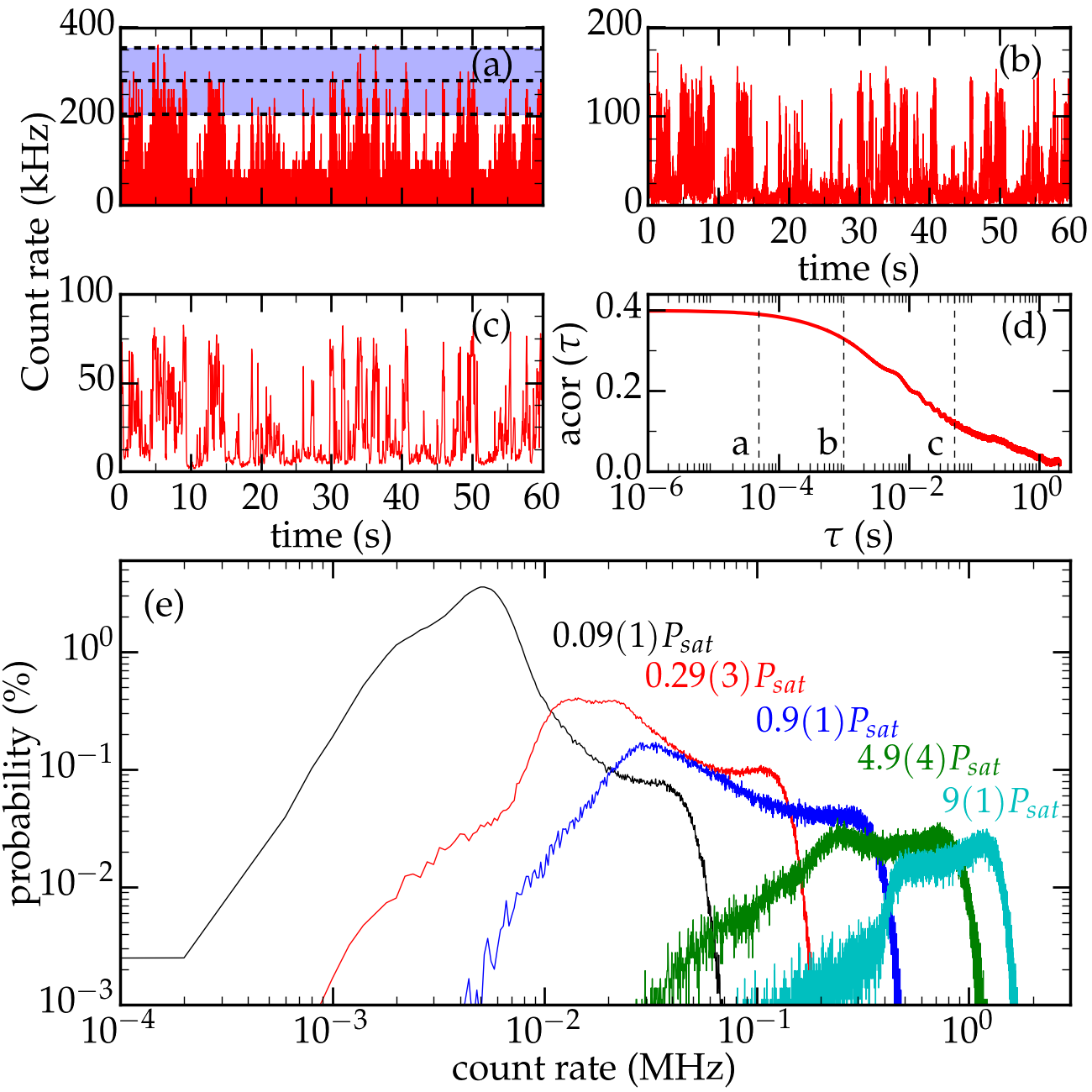} \caption{Time trace of the RF signal for constant $\lambda$ and $V_{g}$ with $P_{exc}=0.029(3)P_{sat}$. The shaded area in (a) represents the expected number of photons for an static TLS with a shot noise contribution. The arrival time of the photons were binned using (a) $T_{bin}=50$~$\mu$s, (b) $T_{bin}=1$~ms and (c) $T_{bin}=50$~ms. (d) Autocorrelation function obtained from the time trace with $T_{bin}=1$~$\mu$s. The dashed lines mark the time bin of the time traces. (e) Empirical probability of counting $I_{tot}$ photons in a time interval $T_{bin}=5$~ms for $P_{exc}=0.09(1)P_{sat}$ (black); $P_{exc}=0.29(3)P_{sat}$ (red); $P_{exc}=0.9(1)P_{sat}$ (blue); $P_{exc}=4.9(4)P_{sat}$ (green);
$P_{exc}=9(1)P_{sat}$ (cyan).}
\label{fig2} 
\end{figure}

The experimental saturation curve is altered because the integration time of $50$~ms in these measurements is long relative to the spectral fluctuation dynamics. This is demonstrated in the time trace measurements acquired by setting $\lambda$ and $V_{g}$ constant and recording the arrival time of the photons in a time tagging mode, shown in Fig.~\ref{fig2}~(a-c). In Fig.~\ref{fig2}~(a), the shaded area represents the expected number of photons extracted from an ideal static TLS with the shot noise contribution. The autocorrelation function $acor(\tau)$ of the time traces presented in Fig.~\ref{fig2}~(a-c) reveals that the spectral fluctuations have dynamics on the millisecond time-scale, as shown in Fig.~\ref{fig2}~(d). This suggests that the environment of the QD may be considered static in a time interval smaller than $50$ $\mu$s, reinforcing previous results \citep{Kuhlmann_NatPhys_2013,Matthiesen_SciRep_2014,prechtel_PRX_3,hansom_APL_105}. These dynamics confirm that the probable source of spectral fluctuations is a dynamic charge density in the sample. We note that other samples of the identical design grown in different growth chambers do not show these dynamics (e.g. behaviour similar to the static TLS is obtained) and therefore conclude that impurities during the growth are the source of the intrinsic charge noise in the sample. The empirical probability of collecting $I_{tot}$ photons, displayed in Fig.~\ref{fig2}~(e), is obtained through normalized histogram of the time traces. They reveal that the charge noise is present even for excitation powers greater than $P_{sat}$ (static case). In this case, the power broadening contributes to the decreasing visibility of the off-resonance distribution.

\subsection{Power spectrum}

\begin{figure}[hb]
\centering \includegraphics[width=1\linewidth]{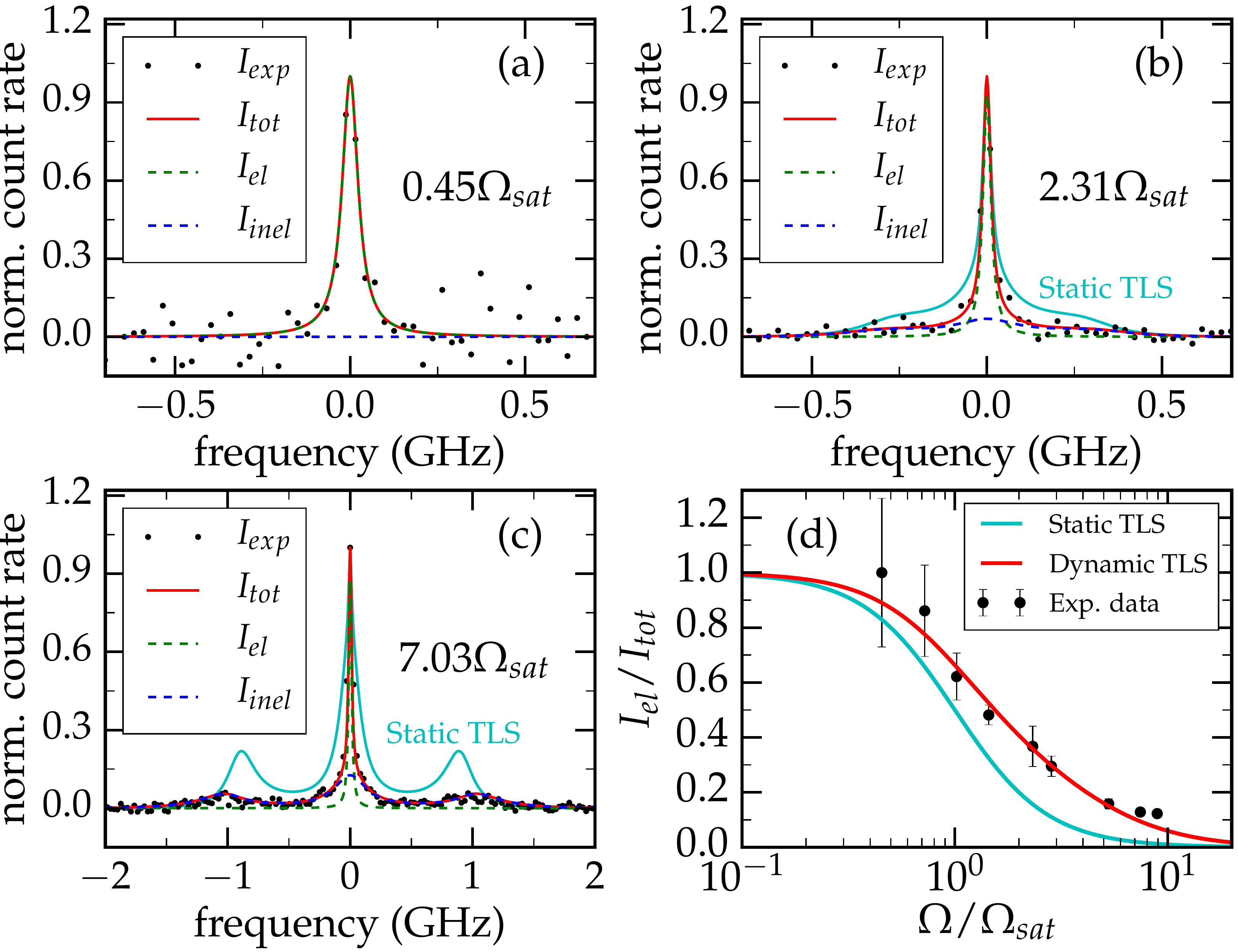} \caption{Power spectrum measurement (black points) fitted with Lorentzian peaks (solid red line), accounting for the elastic scattering (dashed green) and the inelastic scattering (dashed blue) contributions. The solid cyan line (b-c) represents the power spectrum for an ideal static TLS. The Rabi frequencies are $0.45\Omega_{sat}$ (a), $2.31\Omega_{sat}$ (b) and $7.03\Omega_{sat}$ (c). (d) The ratio between the number of elastically scattered photons and the total number of photons (black points) fitted as a TLS under spectral fluctuation (solid red) using
$w=2.4(6)$ $\mu$eV. The solid cyan line represents the ratio $I_{el}/I_{tot}$ for an ideal static TLS.}
\label{fig3} 
\end{figure}

We now spectrally characterize the resonantly scattered photons from the QD in the noisy environment. Fig.~\ref{fig3}~(a-c) shows high resolution ($27$~MHz) spectra from the X$^{1-}$ transition recorded using a scanning Fabry-Perot interferometer for excitation powers below, near, and above saturation ($\Omega=0.45\Omega_{sat}$, $\Omega=2.31\Omega_{sat}$, and $\Omega=7.03\Omega_{sat}$, respectively). Here, we apply modest magnetic field ($B_{z}^{ext}=1$~T) in the growth direction (Faraday geometry) to screen the effect of nuclear spin fluctuations \citep{Ralph_PRL_2016}, and excite the highest energy transition. Below saturation, the power spectrum is dominated by elastically scattered photons, as expected for a two-level system with negligible pure dephasing. As the Rabi frequency is increased, an inelastic peak with $\Gamma/2\pi\approx255$~MHz gains in intensity. Finally, above saturation, the Mollow triplet
consisting of two side-bands (split from the center inelastic peak by the Rabi frequency) as well as the elastic peak are observed. With increasing excitation power, we observe increasing deviation from ideal two-level (static TLS) behaviour due to the spectral fluctuations. The side bands of the Mollow triplet are slightly shifted to higher frequencies and strongly suppressed when compared with the elastic peak. We fit the data by numerically integrating the power spectrum over detuning with a normal distribution as the probability density function (PDF) (Fig.~\ref{fig3}~(b-c)), which is a good description for the electric field probability distribution under the assumption of many contributing charge fluctuations observed over a long time period \citep{Matthiesen_SciRep_2014}. Fig.~\ref{fig3}~(d) shows
the discrepancy between the measured $I_{el}/I_{tot}$ (black circles) and the expected curve for a static TLS (solid cyan), which is corrected with the inclusion of the spectral fluctuations through numerical integration over the detuning with a normal distribution of width $w=2.4(6)$~$\mu$eV as PDF (solid red). We observe that the spectral fluctuations increase the ratio $I_{el}/I_{tot}$, in agreement with previous results \citep{Konthasinghe_PRB_2012}. Note that the phonon
sidebands - which reduce $I_{el}/I_{tot}$\citep{Smith_arxiv_2016} - are not observed in the high-resolution spectra and therefore not accounted for in Fig.~\ref{fig3}~(d). The width of the PDF of the charge noise is below the rough estimate from the five successive detuning spectra in Fig.~\ref{fig1}~(b) ($\sim10$~$\mu$eV). The origin of this disparity is that photon scattering is limited by the width $L$ of the QD transition (including power broadening
effects) while the power spectrum is obtained with a constant $\lambda$ and $V_{g}$. Consequently, when the detuning is greater than $L$, photons are not collected in great quantity and this scenario does not have significant contribution in the power spectrum. We note that the contribution to the RF signal from the lowest energy X$^{1-}$ transition of the Faraday geometry is negligible due to the large energy separation ($\sim107$ $\mu$eV) compared to spectral fluctuations.

\subsection{Two-photon interference}

\begin{figure}[hb]
\centering \includegraphics[width=1\linewidth]{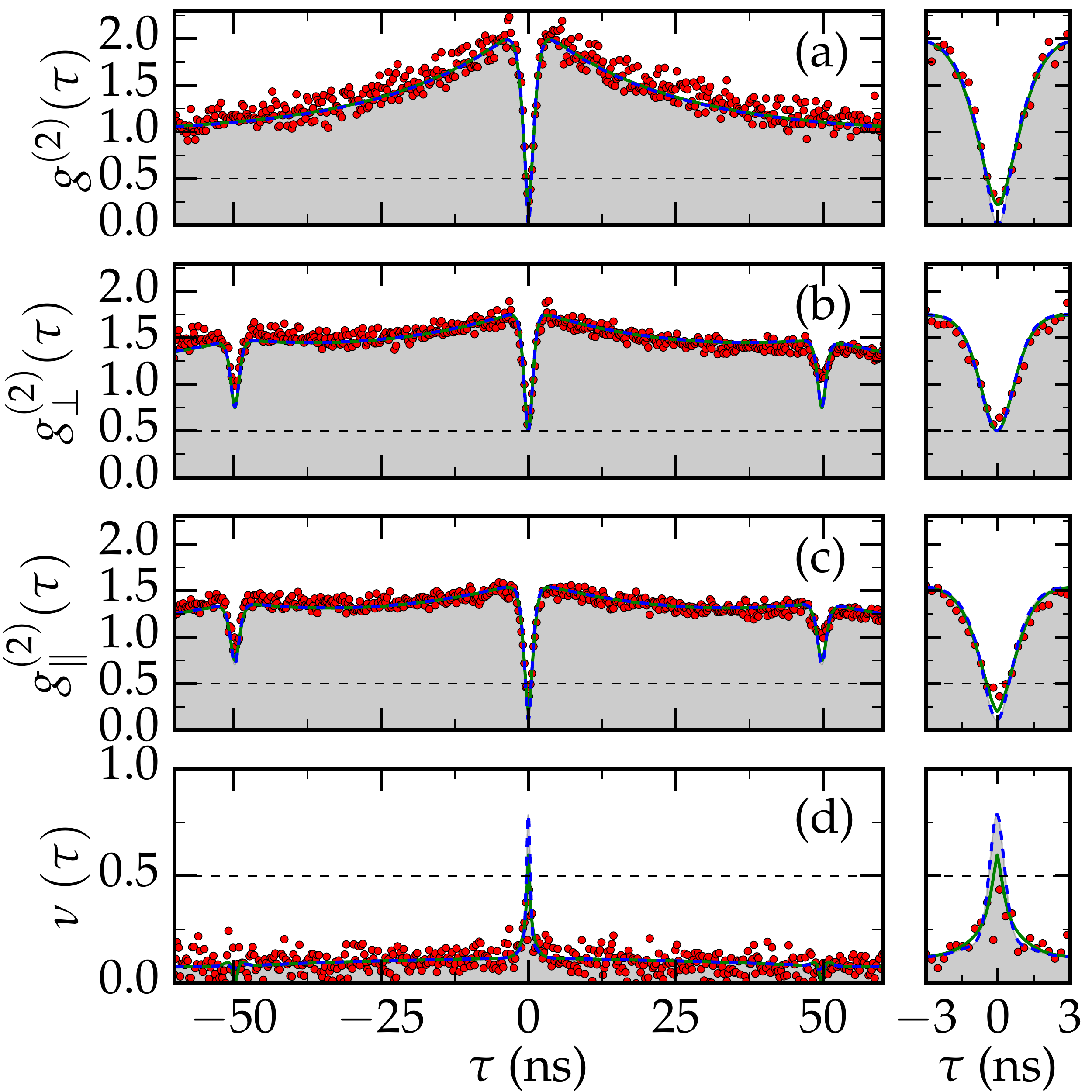} \caption{(a) $g^{(2)}(\tau)$ measurement (red circles) with fitted $g^{(2)}(0)=22\%$ (solid green) and deconvolved $g^{(2)}(0)=0.1\%$ (dashed blue). (b-c) Two-photon interference measurement with the photon polarization at the beam splitter orthogonal (b) or parallel (c) to each other (red circles) with fitted $g_{\perp}^{(2)}(0)=51\%$ and $g_{\parallel}^{(2)}(0)=20\%$ (solid green), and deconvolved $g_{\perp}^{(2)}(0)=50\%$ and $g_{\parallel}^{(2)}(0)=11\%$ (dashed blue). (d) Visibility of the two-photon interference (red circles) with fitted $\nu(0)=60\%$ (solid green) and deconvolved $\nu(0)=79\%$ (dashed blue).}
\label{fig4} 
\end{figure}

Second-order correlation experiments confirm the single photon nature of the RF signal, as shown in Fig.~\ref{fig4}~(a). Here significant bunching around time zero is observed due to optical spin pumping, a consequence of the non-zero external magnetic field \citep{Ralph_PRL_2016,Xu_PRL_2007,Hansom_NatPhys_2014,Dreiser_PRB_2008}. We use a fitting procedure similar to Malein et al \citep{Ralph_PRL_2016}.
Deconvolution of the fit with the response function of the detectors and electronics yields (dashed blue line) yields $g^{(2)}(0)=0.1\%$ with $\Omega=0.39\Omega_{sat}$.

In Fig.~\ref{fig4}~(b) and (c) we can observe the two-photon interference measurements when the polarization of the photons are orthogonal ($g_{\perp}^{(2)}$)
and parallel ($g_{\parallel}^{(2)}$) to each other, respectively, for $\Omega=0.45\Omega_{sat}$. For the distinguishable case, the fit (solid green) and deconvolved curve (dashed blue) give $g_{\perp}^{(2)}(0)$ equal to $51\%$ and $50\%$, respectively. When the polarizations are aligned in parallel, the fit gives $g_{\parallel}^{(2)}(0)=20\%$, while the minimum value of the deconvolved curve is $11\%$. The visibility $\nu(\tau)=[g_{\perp}^{(2)}(\tau)-g_{\parallel}^{(2)}(\tau)]/g_{\perp}^{(2)}(\tau)$ is shown in Fig.~\ref{fig4}~(d) and in the inset (right side panel) with the maximum value of the raw fit and deconvolved curves being $60\%$ and $79\%$, respectively. Low-resolution spectra reveal that $\sim10\%$ of the power spectrum is due to photons scattered into
phonon sidebands. These incoherently scattered photons, not filtered in the two-photon interference experiment, are expected to reduce the maximum possible visibility to $\sim80\%$. Therefore, we conclude the strong spectral fluctuations due to charge noise do not degrade the indistinguishability of successively emitted photons within the $50$ ns time delay used in the unbalenced MZ interferometer.

\section{Summary}

We have presented the photon emission statistics from a quantum dot undergoing strong spectral fluctuations due to a large number of charge traps in its environment. In our device, the fluctuating charges - with dynamics in the millisecond timescale - cause a Stark shift up to $\sim10$ $\mu$eV even with resonant excitation. This leads to flickering in intensity which decreases the time-averaged count rates for excitation powers below saturation and, consequently, the signal
to background ratio. Compared to an ideal, static two-level system, the RF power spectra exhibit shifted and inhibited Rabi side-bands. Importantly, the central peak experiences negligible inhomogeneous broadening due to the spectral fluctuations. Rather the ratio of elastic to inelastic scattered light is affected by the charge noise. Finally, two-photon interference experiments show that photons generated within a short time delay ($50$ ns) are highly coherent and indistinguishable. We conclude that, in-spite of the flickering intensity due to spectral fluctuations, the specificity of resonance florescence enables the
generation of highly coherent and indistinguishable photons from two-level quantum emitters in a noisy environment.

\section{Acknowledgements}

This work was supported by the EPSRC (Grants EP/I023186/1, EP/K015338/1, EP/M013472/1 and EP/G03673X/1) and by ERC Starting Grant No. 307392. B. D. G. thanks the Royal Society for a University Research Fellowship.

\section{References}

 \bibliographystyle{h-physrev}

\end{document}